\pdfoutput=1
\documentclass{PoS}

\usepackage{amssymb,amsmath}

\title{Universal Aspects of Deconfinement: Interfaces, Flux Tubes and Self-Duality in 2+1 Dimensions}

\ShortTitle{Universal Aspects of Deconfinement}

\author{\speaker{Lorenz von Smekal}\\ 
        Institut f\"ur Kernphysik, Technische Universit\"at Darmstadt, D-64289 Darmstadt, Germany\\
        E-mail: \email{lorenz.smekal@physik.tu-darmstadt.de}}

\author{Sam R.~Edwards\\
        Institut f\"ur Kernphysik, Technische Universit\"at Darmstadt, D-64289 Darmstadt, Germany\\
        E-mail: \email{edwards@crunch.ikp.physik.tu-darmstadt.de}}

\author{Nils Strodthoff\\
        Institut f\"ur Kernphysik, Technische Universit\"at Darmstadt, D-64289 Darmstadt, Germany\\
        E-mail: \email{nstrodt@crunch.ikp.physik.tu-darmstadt.de}}

\abstract{We study center vortex free energies and 't Hooft's electric
  fluxes on the lattice in 2+1 dimensions, where $SU(2)$ for
  example, is in the universality class of the 2$d$ Ising model. This
  places a wealth of exact results at our fingertips. In particular,
  spacelike center vortices in $SU(2)$ near criticality correspond to
  spin interfaces in the 2$d$ Ising model, whose universal scaling
  functions are known exactly. We exploit this to locate the
  deconfinement transition with unprecedented precision and
  subsequently for a finite size scaling analysis, where the
  self-duality of the $2d$ spin model is reflected in a duality
  between the spacelike vortices and confining electric fluxes. The
  corresponding relation between the string tension and its dual in
  the high temperature phase is arguably the simplest example of a
  universal amplitude ratio. Around the transition, both can be
  efficiently extracted from the exact results with a global
  one-parameter fit which allows straightforward continuum
  extrapolation.} 

\FullConference{The XXVIII International Symposium on Lattice Field
  Theory, Lattice2010\\ 
		June 14-19, 2010\\
		Villasimius, Sardinia, Italy}

\begin{document}

\section{Introduction}

\vspace{-.2cm}

The finite temperature deconfinement transition in pure $SU(N)$ gauge
theories in $d+1$ dimensions is very well understood in terms of the
spontaneous breakdown of their global $Z_N$ center symmetry. This
symmetry is faithfully represented by the fundamental Polakov loops
which live on $d$ dimensions and develop a non-zero expectation value
in the deconfined broken phase much like the spontaneous magnetization
in a $d$ dimensional $q$-state Potts model \cite{Wu:1982ra} with $q =
N$.   

In order to be able to apply the powerful tools of universality and
scaling near a critical point, here we are particularly interested in
cases where this transition is of $2^\mathrm{nd}$ order. For pure QCD
with $N=3$ colors in $3+1$ dimensions this is of course not the
case. The transition is first order, but only just. If we either
reduce the number of colors to $N=2$ or the dimensions to $2+1$, 
we obtain Yang-Mills theories with a second order deconfinement
transition within the universality class of $q$-state Potts models
with $q=2$ (Ising) in $d=3$ dimensions or with $q=2$ and $3$ in
$d=2$. The $q=4$ Potts model in 2 dimensions is interesting because it
is known  \cite{Baxter:2000ez} to have a $2^\mathrm{nd}$ order
transition {\em and} to fall precisely on the separatrix $q_c(d)$ with
respect to the order of the transition in the $(q,d)$-plane, where
only $1^\mathrm{st} $ order transitions occur for $q> q_c(d) $, {\it
  i.e.}, $q_c(2)=4$. The corresponding $SU(4)$ gauge theory in 2+1
dimensions has recently been studied for example in
\cite{deForcrand:2003wa,Holland:2007ar,Liddle:2008kk}. The conclusion
in \cite{Holland:2007ar} is that the transition is weakly
$1^\mathrm{st}$ order, unlike the $q=4$ Potts case. It nevertheless
seems that there is a rather wide range of intermediate length scales
where at least approximate Potts scaling can be observed. One might
then like to understand, for example, why among the wider class of
Ashkin-Teller models it is the standard $q=4$ Potts scaling that is
relevant here, and whether this can be derived from an effective
Polyakov-loop model.  
   
We will present more results specific for $SU(3)$ and $SU(4)$ in 2+1
dimensions in \cite{NilsLat2010}. One general property of the 2$d$ Potts
models is that they are self-dual for all $q$. We will demonstrate
below, how this self-duality is reflected in a duality between the
spacelike center vortices and the confining electric fluxes of the
gauge theory. Herein we will mainly report on numerical results for
$SU(2)$ in 2+1 dimensions for which many exact results
from the 2$d$ Ising model are available.    

\vspace{-.4cm}

\section{Center Vortices and Interfaces}

\vspace{-.2cm}

For pure $SU(N)$ gauge theories in a finite Euclidean box, 't Hooft's
twisted boundary conditions fix the total $Z_N$-valued center flux
through each plane of the box, {\it i.e.}, the total number modulo $N$ 
of center vortices piercing that plane \cite{'tHooft:1979uj}. 
At finite temperature $T$, the differences in free energy between
these twisted ensembles and the periodic one define the corresponding
center-vortex free energies. If they tend to zero in the thermodynamic
limit, then these vortices condense which leads to an area law for those
Wilson loops that feel their disordering phases in the corresponding
plane of the twist.  
In an $L^d \times 1/T$ box, we have to distinguish between temporal
and magnetic twist. The latter is defined in purely spatial planes and
corresponds to the  $Z_N$-valued magnetic flux $\vec m$ of static center
monopoles. Their free energy tends to zero with $L\to\infty$ as $\exp\{
-\sigma_s(T) L^2 \}$ at all $T$ \cite{vonSmekal:2002gg}, corresponding
to the area law for spatial Wilson loops with spatial string tension
$\sigma_s(T)$.   

\begin{figure}[t]
\vspace*{-.2cm}
\centerline{\includegraphics[width=0.8\linewidth]{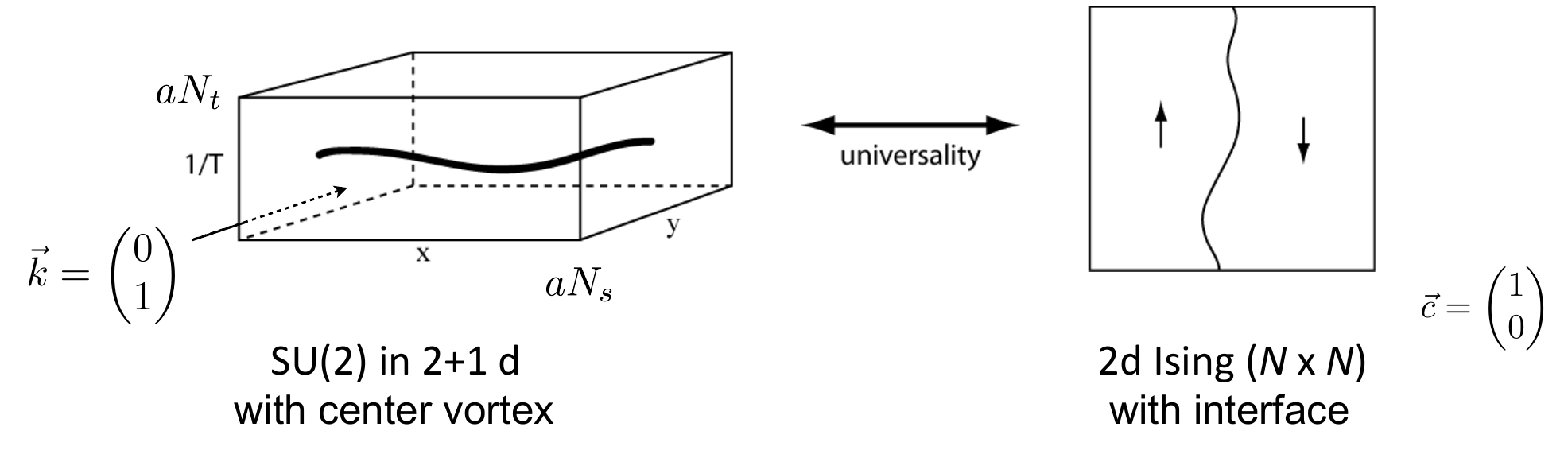}}
\vspace{-.3cm}
\caption{Spatial center vortices share their universal behavior with
  interfaces in a spin model. \vspace*{-.6cm}} 
\label{fig1}
\end{figure}

Relevant to the deconfinement transition, however, are only the
temporal twists. These are characterized by $d$-dimensional vectors
$\vec k$ of integers $\bmod\ N$, {\it i.e.}, with components 
$k_i \in Z_N$ representing the twist in the temporal plane of orientation
$(0,i)$, with total center flux $\exp( 2\pi i\,
k_i/N)$ through that plane. See Fig.~\ref{fig1} for an illustration of
such a vortex ensemble in $SU(2)$. We denote the 
partition functions of the various ensembles with temporally twisted boundary
conditions by $Z_k(\vec k)$. The corresponding center-vortex free
energies per $T$ are then given by  $F_k(\vec k) = - \ln(Z_k(\vec k)/Z_k(0)) $,
where $Z_k(0)$ stands for the periodic ensemble. In the vicinity of a
second order deconfinement transition these vortex free energies show 
the universal behavior of interfaces in the respective $d$
dimensional Potts model. These interface free energies, $F_I(\vec c) = -
\ln(Z_q(\vec c)/Z_q(0))$,  are likewise
obtained from ratios of partition functions $Z_q(\vec c) $ with
boundary conditions denoted by a $d$ dimensional vector $\vec c$ such
that the $q$ spin states $s_{\vec x} = 0,\, 1,\,  \dots q-1  $ are
cyclically shifted across the boundary, {\it i.e.}, $s_{\vec x + \vec
  e_i L} = s_{\vec x} + c_i \bmod q$, with $c_i =  0,\, 1,\,  \dots
q-1$.  For $q=2$, the Ising model, this simply means anti-periodic in
direction $i$, if $c_i = 1$, and periodic otherwise.  

Interfaces(vortices) are suppressed in the low(high) temperature
ordered(deconfinement) phase below(above) $T_c $. Here, the ratios
$R_q(\vec c) \equiv  Z_q(\vec c)/Z_q(0)$ and $R_k(\vec k) \equiv
Z_k(\vec k)/Z_k(0)$ both tend to zero as we approach the thermodynamic limit.  
Complementary to that, in the disordered(confined) phase, it is
the interface(vortex) free energies that tend to zero such that the
ratios $R_q$, $R_k $ approach 1  for all boundary conditions. Only at the
critical temperature  $T=T_c $ do these ratios converge to non-trivial
and universal fixed points. In 2 dimensions these universal numbers  
$0< R_c^{(m,n)} < 1$ have been obtained exactly  \cite{PhysRevB.38.565}, in
terms of Jacobi theta functions, for all cyclic boundary conditions $\vec c =
(m,n)$ and all Potts models with $2^\mathrm{nd}$ order transitions,
{\it i.e.}, for $q=2$, $3$ and $4$.

These universal ratios can be used to determine the critical couplings
$\beta_c$ of the transition in the 2+1 dimensional gauge theories with
high precision \cite{Edwards:2009qw}, by requiring that $\beta =
\beta_c$ at $R_k = R_c$. For $SU(2)$, 
where,
\vspace{-.2cm}
\begin{equation}
   R_c^{(1,0)} = 1/(2^{3/4}+1)\, , \;\; \mbox{and} \;\; 
R_c^{(1,1)} = (2^{3/4}-1)/(2^{3/4}+1)\, ,
\end{equation}
from the 2$d$ square Ising model, for example, this allowed us to determine the 
critical couplings 
for lattices with up to $N_t = 9$ links in the Euclidean time
direction with a typically by two orders of magnitude increased precision
where there were literature values available
\cite{Teper:1993gp,Engels:1996dz,Liddle:2008kk}.  
The high precision allowed us to reliably determine the
subleading $1/N_t$-corrections to the linearly increasing behavior of $\beta_c$
with $N_t$ near the continuum limit \cite{Edwards:2009qw},
\vspace{-.2cm}
\begin{equation}
\beta_c(N_t) = 
1.5028(21)\, N_{t}+0.705(21)-0.718(49)\, \frac{1}{N_{t}} \, .   
\end{equation}  
The slope of the leading term determines the critical temperature
$T_c$ in units of the dimensionful continuum coupling $g_3^2 $ of the
(2+1)$d$ theory, yielding $T_c/g_3^2 = 0.3757(5)$. By standard
arguments, from this result one can also read off the temperature
dependence of the coupling at a given $N_t$,
\begin{equation} 
\beta(t) - \beta_c = 4N_t \, \frac{T_c}{g_3^2} \, t -  \frac{0.270(2)}{N_t}
\, \frac{g_3^2}{T_c} \, \frac{t}{1+t} \, + \, \mathcal O(1/N_t^2) \; , \quad t =
\frac{T}{T_c} - 1 \; . \label{tbeta}
\end{equation}
For each fixed $N_t$, we thus have precise control of the temperature
by varying the lattice coupling $\beta$. The physical length of the
spatial volume then follows from $L = a N_s = T_c^{-1} (N_s/N_t)/(1+t)$.

\begin{floatingfigure}[r]
\includegraphics[width=0.4\linewidth]{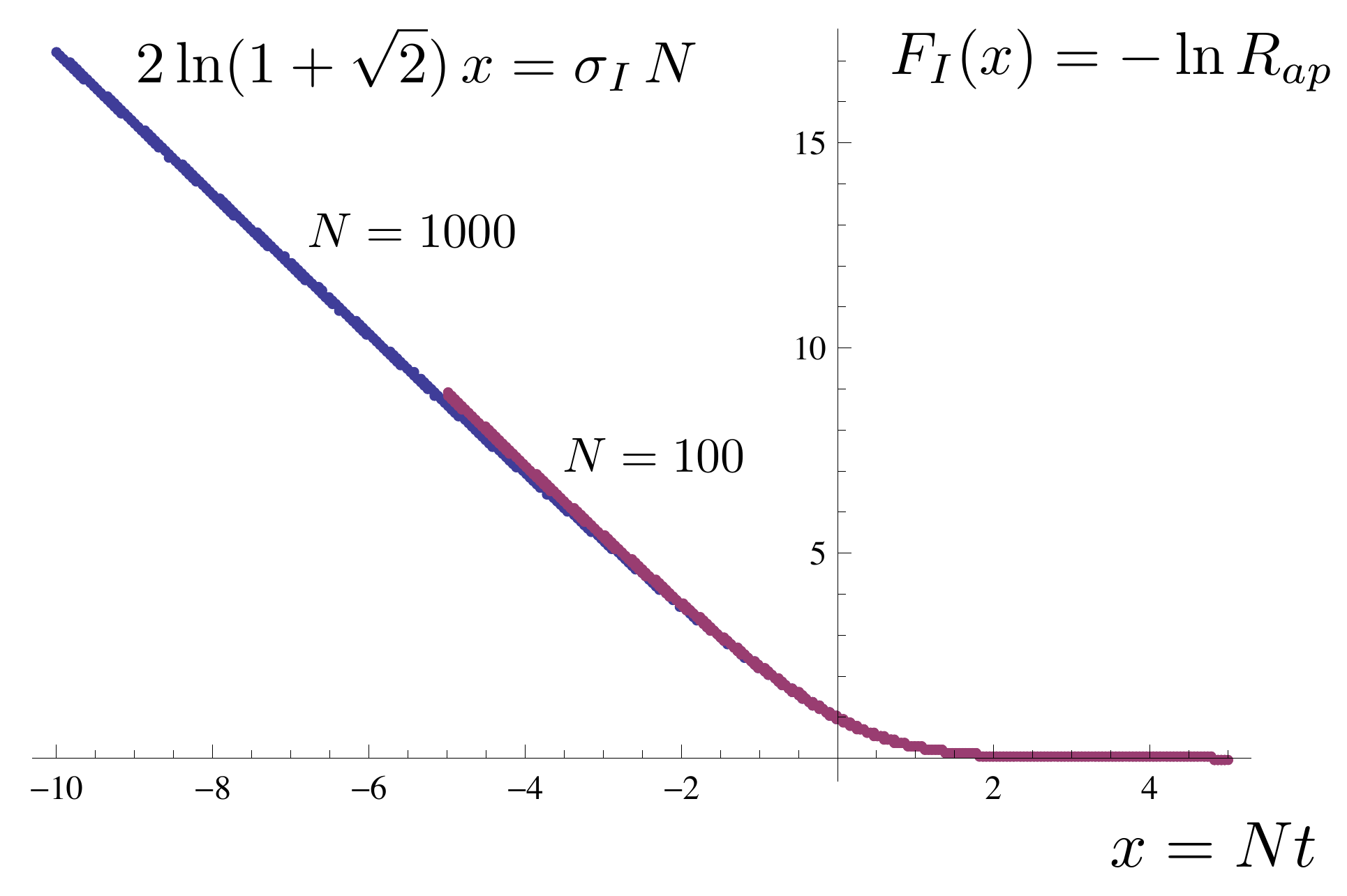}
\vspace{-.4cm}
\caption{Universal scaling function $F_I(x)$ 
 from the exact finite-volume
 partition functions of the $N\times N$
  square  Ising model in \cite{1999JPhA.32.4897W}.
\vspace{-.6cm}
} 
\label{fig2}
\end{floatingfigure}

With these we can perform a finite-size-scaling
analysis near the $2^\mathrm{nd}$ order phase transition where
generalized couplings such as the vortex-ensemble ratios $R_k$, for
sufficiently large $L$, only depend on $L^{1/\nu} t$, with the
correlation-length critical exponent $\nu $, in a universal way. In
particular, for (2+1)$d$ $SU(2)$, this dependence is determined by that
of the corresponding ratios $R_{q=2}$ of finite-volume partition functions 
of the 2$d$ square Ising model, with $\nu =1$, which have been
obtained exactly for all combinations of periodic and anti-periodic
boundary conditions in \cite{1999JPhA.32.4897W}.   

For one antiperiodic direction, the corresponding exact interface free
energies per temperature $F_I = - \ln R_{ap}$, where $R_{ap} $
stands for $R_{q=2}(\vec c)$ with either $\vec c = (1,0) $ or $ (0,1)$
which are the same on an $ N\times N$ square lattice, are plotted over the
finite-size-scaling variable $x \equiv Nt$ for $N=100$ and $N=1000$ in
Fig.~\ref{fig2}.  The asymptotic slope in the ordered phase ($t<0$)
for large $N$ approaches $2\ln(1+\sqrt 2)$ which follows from
Onsager's famous result \cite{Onsager:1943}
for the tension of a straight interface in the
infinite volume limit, $\sigma_I = 2 K + \ln\tanh K $ with
spin-coupling per temperature $K=J/T$, near criticality at $K_c =
\ln(1+\sqrt{2})/2$. Deviations from that,
{\it e.g.}, observed for $N=100$ around $x=-5$, are
due to subleading (non-universal) finite-volume effects.      

We have calculated the $SU(2)$ vortex-ensemble ratios $R_k(\vec k)$
with the algorithm of \cite{deForcrand:2000fi} as previously for the
(3+1)$d$ theory in \cite{deForcrand:2001nd}, here in (2+1)$d$ 
for $N_t = 2$ to $10$, each with spatial sizes up to  $N_s=96$  and in
a suitable window of $\beta$'s around $\beta_c$ to test the
finite-size scaling (FSS). For each fixed $N_t$ we generally observe 
very good scaling of the available data for all $N_s$ when plotted
over the FSS variable $x \equiv L T_c t = (N_s/N_t) \,
t(\beta)/(1+t(\beta)) $ (assuming $\nu =1$ for the 2$d$ Ising 
universality class) with $t(\beta)$ from Eq.~(\ref{tbeta}). For the
resulting vortex free energies $F_k$ (for $\vec k = (1,0) $ and $(1,1)$)  
we then obtain accurate one-parameter fits via $F_k(x) = F_I(-\lambda
x) $ to the exact universal scaling functions $F_I(x)$ (for $(1,0)$
and $(1,1)$ b.c.'s). This determines the single non-universal
parameter $\lambda $ which is dimensionless and relates 
the $SU(2)$ FSS variable to the Ising one (whereby the minus sign
reflects the interchange of high and low temperature phases between
the two).       

\vspace{-.4cm}
  
\section{Self-Duality in 2+1 Dimensions}

\vspace{-.2cm}

\begin{figure}[t]
\includegraphics[width=\linewidth,trim=.8cm 0 0
0,clip=true]{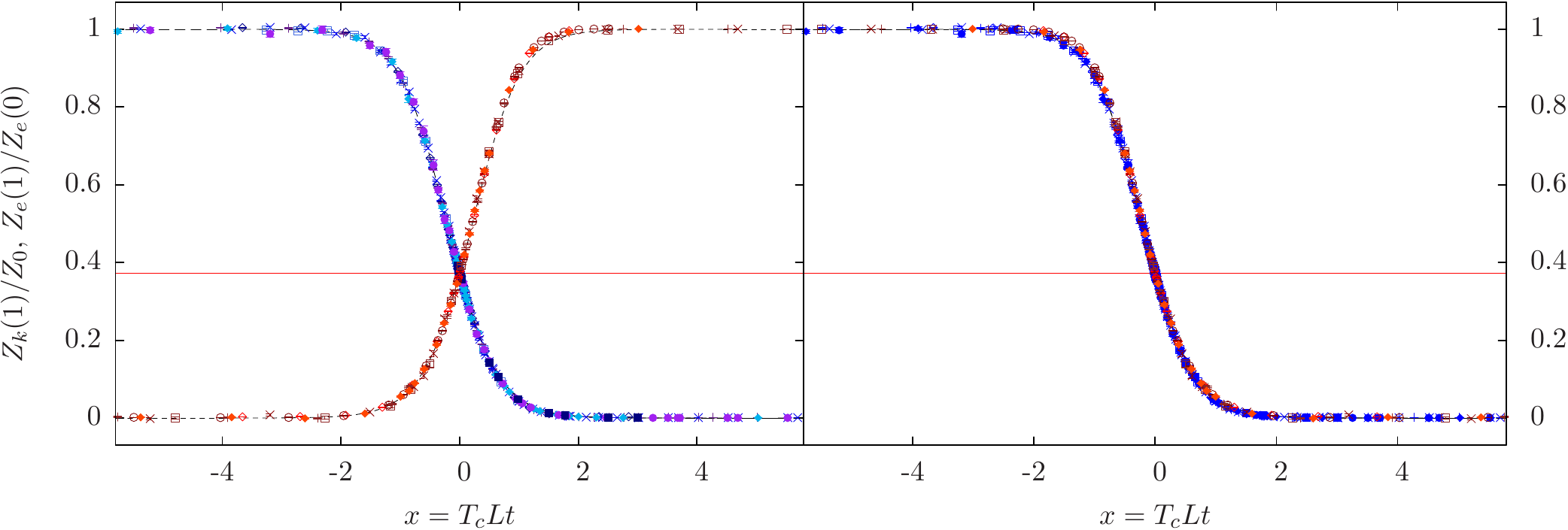} 

\vspace*{-5cm}
\leftline{\footnotesize\hskip .8cm $R_k(x) $ \hskip 4.9cm
  $R_e(x) $ \hskip .4cm $R_k(x),\, R_e(-x) $}
\vspace{4.2cm}
\caption{The ratio  $R_k$ of the partition function with twist  $\vec k =
  (1,0)$ over the periodic ensemble compared to that of one unit of
  electric flux relative to the no-flux ensemble (left), $R_e =
  Z_e(\vec e)/Z_e(0)$ with $\vec e =   (1,0)$, 
  and to its mirror image (right). The data here is obtained for $N_t
= 4$ and various spatial volumes up to $N_s=96$. \vspace{-.4cm}}
\label{fig3}
\end{figure}

The 2$d$ $q$-state Potts models are self-dual for all $q$. This has
long been known for infinite systems. A particularly simple proof
based on the random bond-cluster representation was given in
\cite{Wu:1982ra}. Far less is known in a 
finite volume with translationally invariant boundary conditions,
however, where ensembles with different boundary conditions 
mix under duality transformations. One case where this is known is the
3$d$ Ising/$Z_2$-gauge theory system  \cite{Caselle:2001im}. For the
Potts model, using the random cluster methods developed in
\cite{PhysRevB.38.565}, we were able to obtain the following exact
duality transformations for all $q$ on a finite 2$d$ torus as discrete
2$d$ Fourier transforms over all ensembles $Z_q^{(m,n)}$ with
cyclically shifted boundary conditions $\vec c = (m,n)$,  
\vspace{-.3cm}
\begin{equation} 
Z_q^{(-s,r)}(\widetilde K) \,=\, \Big( \, \frac{e^{\widetilde K}
  -1}{e^K-1} \,\Big)^{N_\mathrm{sites}} \frac{1}{q} \, \sum_{m,\, n}
\, e^{\frac{2\pi i}{q} (rm+sn)} \; Z_q^{(m,n)}(K) \; , \quad m,\, n,\,
r,\, s \,= \, 0, 1, \dots q-1\; , \label{Duality} 
\end{equation}
%
where $N_\mathrm{sites}$ is the total number of sites of the graph on
the 2$d$ torus ({\it i.e.}, 
$N^2$ on an 
$N\times N$ square lattice), $K=J/T$ the coupling per temperature,
and $\widetilde K$ its dual obtained from $(e^{\widetilde K}-1)(e^K
-1) = q$, with temperature mirrored around criticality at $K=
\widetilde K = K_c = \ln(1+\sqrt q)$.  

For $q=2$, and with $K \to K/2$, conventionally, this relation reduces
to an analogous known result for the Ising model \cite{Muenster}. For
$q=3$ it agrees with a corresponding result \cite{Bugrii:1997ry}
obtained for the planar  or vector Potts model which is equivalent to
the standard one in that case. The proof of the general formula in
Eq.~(\ref{Duality}), for all $q$-state Potts models on a 2$d$ torus, will
be given elsewhere \cite{NilsLat2010}. 

Here we only note that the pattern in the finite-volume duality
transforms, by the discrete Fourier transform over the ensembles with
cyclically shifted boundary conditions, is precisely the same as that
between the twisted vortex ensembles $Z_k(\vec k,\vec m) $
in the $d+1$ dimensional pure $SU(N)$ gauge theory and those with
fixed units of electric (and magnetic) flux $Z_e(\vec e,\vec m)$
\cite{'tHooft:1979uj},
\vspace{-.1cm} 
\begin{equation}
  Z_e(\vec e,\vec m) \,= \, \frac{1}{N^d}  \, \sum_{\vec k\in Z_N^d}
  \, e^{2\pi i \, \vec e\cdot\vec k/N}\, Z_k(\vec k,\vec m) \; , 
\end{equation}  
\vspace{-.4cm}

\noindent 
obtained from a $d$-dimensional $Z_N$-Fourier transform over all temporally
twisted $\vec k$-b.c.'s. We are not interested in magnetic flux 
and drop the argument $\vec m$ again, in the following. The role of
the electric flux ensembles is best understood in terms of the 
translationally invariant flux between a fundamental color charge at
some point $\vec x$ in the finite volume and its mirror charge at
$\vec x + \vec e L$ in a neighboring volume in the direction of the
flux $\vec e$ \cite{deForcrand:2001nd}, it contains no ultraviolet
divergent perimeter and no short-distance Coulomb contributions
\cite{deForcrand:2001dp},     
\vspace{-.1cm} 
\begin{equation}
{Z_e(\vec e)}/{Z_e(0)} \, = \, \frac{1}{N}\, \big\langle \mbox{tr}\big( P(\vec x) P^\dagger (\vec
x+ \vec e L) \big) \big\rangle_{\mbox{\scriptsize no-flux}}\, , \label{mirror}
 \end{equation}
\vspace{-.7cm} 

\noindent
where the $P$'s are untraced fundamental Polyakov loops, including
any non-trivial transition functions in the time direction, and the
subscript indicates that their expectation value is taken
in the enlarged no-flux ensemble $Z_e(0) = \sum_{\vec k}  Z_k(\vec k)
/N^d $. 
Without loss, the form in (\ref{mirror}) assumes  periodic spatial
b.c.'s and is manifestly invariant under spatially periodic
gauge transformations.  


Of course, the duality relations provide exact maps between the high
and low temperature phases in the spin models, valid at all
temperatures and beyond universality. We do not have such exact duality
relations for the gauge theories, but near criticality, where $K
\leftrightarrow \widetilde K$ amounts to $x \leftrightarrow -x$,   
we expect that the spin model dualities are reflected in analogously
dual relations between the behavior of the spatial center vortices and 
the confining electric fluxes. This is indeed the case in a rather
wide universal scaling window around $T_c$ and, in particular, for 
2+1$d$ gauge theories the self-duality of the spin models implies that
the electric-flux free energies $F_e(\vec e) $ are mirror images of
the center vortex free energies, {\it i.e.}, $F_e(x) = F_k(-x) $, as
demonstrated for $SU(2)$ 
 in Fig.~\ref{fig3}. 

\vspace{-.4cm}

\section{Universal Amplitude Ratios and Continuum Limit}

\vspace{-.2cm}

The interface tension $\sigma_I$ in the ordered phase of the spin
model corresponds to the dual string tension $\tilde\sigma $  
for spatial center-vortex sheets which provides the dual area law
for the spatial 't Hooft loops above $T_c$. Both vanish near the
$2^\mathrm{nd}$ order phase transition, {\it e.g.}, $\tilde\sigma \sim
t^\mu$ for $t\to 0^+$, with a critical exponent that is related to the
correlation length critical exponent $\nu$ via the hyperscaling
relations as $\mu = (d-1) \nu$, {\it i.e.} $\mu=\nu$ in $d=2$ spatial
dimensions. The product of interface tension and correlation length (to
the power $d-1$) is an example of a universal amplitude
ratio \cite{Pelissetto:2000ek}. If we use the exponential correlation
length in the disordered phase of the spin model, $\xi_\mathrm{gap}^+
$, then the universal constant given by  $\sigma_I (\xi_\mathrm{gap}^+)^{d-1}  =
{R_\sigma^+}_\mathrm{gap} $ turns out to be exactly $ 1$ for $q=2,\,
3$ and probably also for $q=4$ in 2 dimensions. In the gauge theory
the exponential correlation length of Polyakov loops in the
confined phase is given by string tension and temperature as
$(\sigma/T)^{-1}$. Therefore, we have a universal relation between
the string tension below and the dual string tension above $T_c$,
\vspace{-.1cm}
\begin{equation}   
 \tilde\sigma \,=\, {R_\sigma^+}_\mathrm{gap} (\sigma/T)^{d-1},  \;\;\;
 \mbox{or} \;\; \sigma = \rho T_c^2 (-t)^\nu +\cdots  
 \;\;\mbox{and} \;\; \tilde\sigma = {R_\sigma^+}_\mathrm{gap} (\rho
 T_c)^{d-1} t^{(d-1)\nu} + \cdots \, ,
\end{equation}
\vspace{-.6cm}

\noindent
where at leading order in the reduced temperature $t$ the only unknown
is the single unique constant $\rho$ that we have introduced here.  
This constant can be extracted from the vortex free energies in the
high temperature phase at asymptotically large $x = T_cL \, t^\nu
$, where $F_k(x) =   {R_\sigma^+}_\mathrm{gap} \rho^{d-1} x^{d-1}$, by
direct numerical simulations. This was done in
\cite{deForcrand:2001nd}. It is numerically very hard and expensive because
larger and larger lattices are needed to stay within the universal
scaling window as $x$ is increased.

\begin{figure}[t]
\vspace*{-.2cm}
\leftline{\includegraphics[width=0.488\linewidth,trim=0.8cm 0
0 0,clip=true]{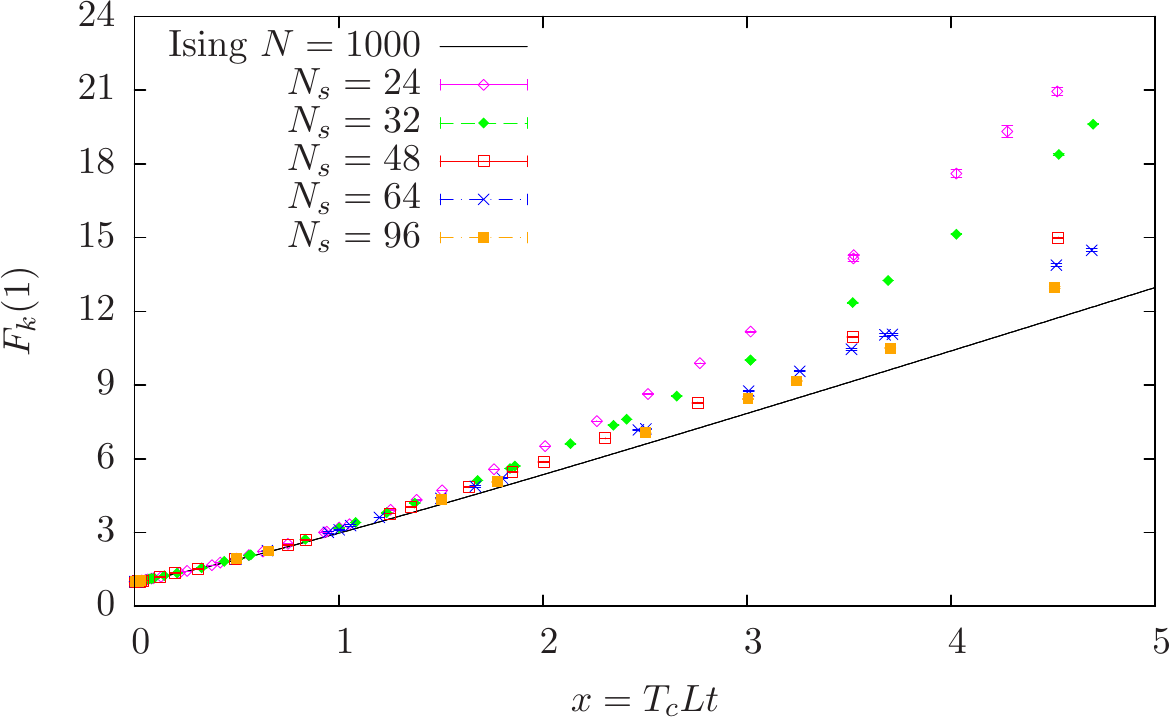} \hfill
\includegraphics[width=0.49\linewidth,trim=0.6cm 0 0 0,clip=true]{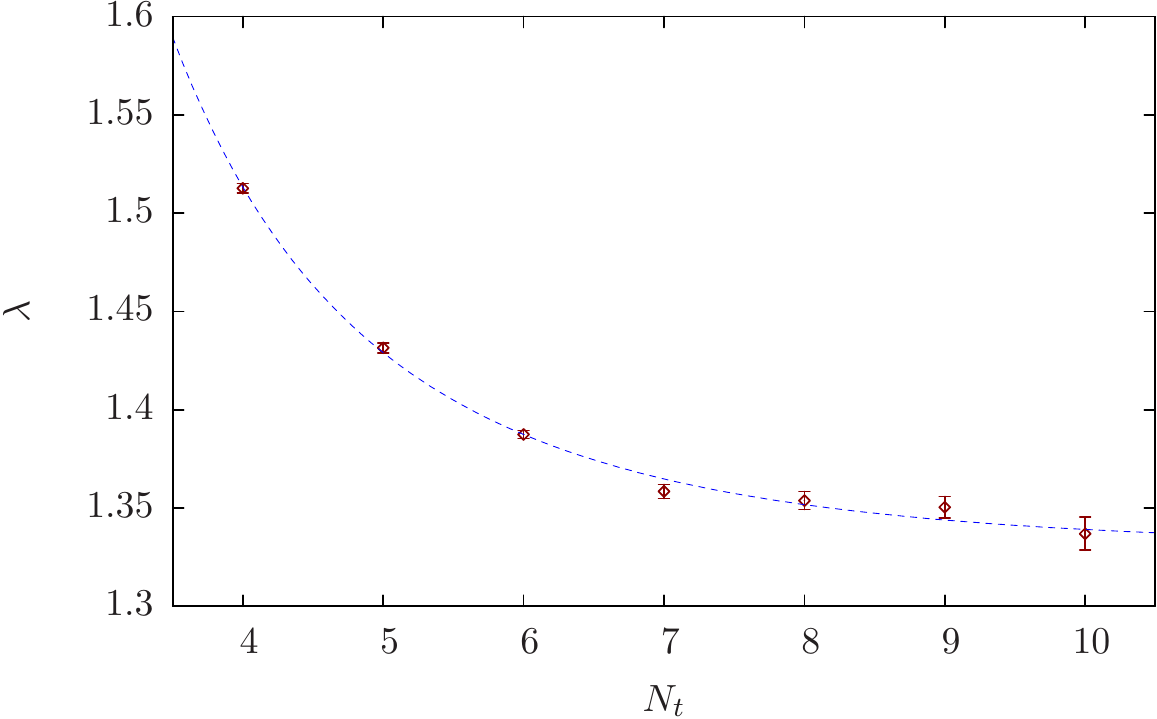}}
\vspace{-4.2cm}
\leftline{\hskip 11.6cm \parbox{4cm}{\scriptsize  The scale parameter $\lambda(N_t)$,\\
                                  fit to $\lambda_\infty + b/N_t + c/N_t^2 $.}}

\vspace{1.2cm}
\leftline{\small\hskip 5cm $\nwarrow $}
\vspace{-.2cm}
\leftline{\scriptsize \hskip 5.4cm $F_I(-\lambda x)\sim \rho x$}

\vspace{1.4cm}
\caption{Vortex free energy $F_k(x)$ in $SU(2)$ above $T_c$ (left), and
   $\lambda(N_t)$ from fits to $F_k(x)=F_I(-\lambda x)$ (right). 
 \vspace{-.6cm}}
\label{fig4}
\end{figure}

This is demonstrated for (2+1)$d$ $SU(2)$ in Fig.~\ref{fig4} where the
universal scaling function is known exactly from the Ising
model. While its linear large-$x$ behavior, $F_k(x) \to \rho x$,
is approached with increasing lattice size, even for the $N_s =96$ lattice 
it is still difficult to extract the asymptotic slope
reliably from the data. But we know the asymptotic slope of
the universal scaling function $F_I(x)$ which is $2\ln(1+\sqrt{2})$. Therefore,
from $F_k(x) = F_I(-\lambda x)$ we obtain $\rho = 2\lambda
\ln(1+\sqrt{2})$ with $\lambda$ from our one-parameter fits
at small $x$ where we have very accurate data.  
Moreover, we have determined this single non-universal parameter for
$N_t = 4,\, 5 \dots 10$ which allows a polynomial fit, 
$ \lambda(N_t) = \lambda_\infty + b/N_t + c/N_t^2 $, with the   
extrapolated result $\lambda_\infty = 1.354(25)$.
This then determines the leading behavior of the continuum 
string tension and its dual around the phase transition,
\begin{equation} 
\sigma = \lambda T_c^2 2\ln(1+\sqrt 2)\, |t| + \cdots , \;\;  t\to 0^-\;,
\;\; \mbox{and} \;\;\; \tilde\sigma = \lambda T_c 2\ln(1+\sqrt 2)\, t
+ \cdots , \;\;  t\to 0^+\; . 
\end{equation} 

\vspace{-.4cm} 

\section{Conclusions}

\vspace{-.2cm} 

Here we presented results for center vortex free energies in the 2+1
dimensional $SU(2)$ gauge theory whose universal properties near the
deconfinement transition can be analyzed in terms of many exact
results from the 2$d$ Ising model. We used these to precisely locate
the transition, to demonstrate finite-size-scaling, and to extract the
continuum string tension and its dual from universal amplitude ratios
together with one-parameter fits to exactly known universal scaling functions.  

We found an exact duality transformation for the $q$-state Potts
models on a finite 2$d$ torus and demonstrated that their self-duality
is reflected in the (2+1)$d$ gauge theory: The free energies of
spatial center vortices are mirror images around $T_c$ of those of the
confining electric fluxes.

More results for $SU(3)$ and $SU(4)$ will be presented elsewhere, see
also \cite{NilsLat2010}. The relevance of center symmetry and center
vortices for confinement in full QCD when including the electromagnetic
interactions of the fractionally charged quarks are
discussed in \cite{SamLat2010}.

\bigskip
\vfill

\noindent\textbf{Acknowledgements:} This work was supported by the
Helmholtz International Center for FAIR within the LOEWE program of
the State of Hesse, the Helmholtz Association Grant VH-NG-332, and the
European Commission, FP7-PEOPLE-2009-RG No.~249203. Simulations were
performed on the high-performance computing facilities of eResearch
SA, South Australia.

\end{document}